\documentclass[aps,preprintnumbers,amsmath,amssymb,showpacs,nofootinbib,twocolumn]{revtex4}
\usepackage{graphicx}
\usepackage{dcolumn}
\usepackage{bm}
\usepackage{epsfig}
\begin{document}
\title{{\boldmath On the $\gamma^*\gamma \to\pi(\eta ,\eta')$} transition form factors}
\author{Dmitri Melikhov$^{1,2,3}$ and Berthold Stech$^4$}
\affiliation{
$^1$HEPHY, Austrian Academy of Sciences, Nikolsdorfergasse 18, A-1050  
Vienna, Austria\\
$^2$Faculty of Physics, University of Vienna, Boltzmanngasse 5, A-1090  
Vienna, Austria\\
$^3$SINP, Moscow State University, 119991 Moscow, Russia\\
$^4$ ITP, Heidelberg University, Philosophenweg 16, D-69120,  
Heidelberg, Germany}
\date{\today}
\begin{abstract}
The surprising results by the BarBar collaboration on the $\pi\gamma$  
transition form factor require new thoughts about the high-$Q^2$ dependence of the form  
factors with virtual photons. We make use of the anomaly sum rule 
[J. Horejsi and O. Teryaev, Z. Phys. C {\bf 65}, 691 (1995).]
which relates the hadron spectral density to the axial anomaly 
[S. Adler, Phys. Rev. {\bf 177}, 2426 (1969); J. S. Bell and R. Jackiw, Nuovo
Cimento A {\bf 60}, 47 (1969).]
We study the quark-hadron duality relation for this sum rule
and find out that the increase of the rescaled form factor $Q^2F_{\pi \gamma}(Q^2)\sim\log(Q^2)$
suggested by the BaBar data requires the presence of a $1/s$-correction term in the relation 
between the one-loop spectral density and the hadron-continuum spectral density.
\end{abstract}
\pacs{11.55.Hx, 12.38.Lg, 03.65.Ge, 14.40.Be}
\maketitle


\section{Motivation and Results}
\begin{figure}[hb!]
\begin{center}
\begin{tabular}{c}
\includegraphics[width=8.4cm]{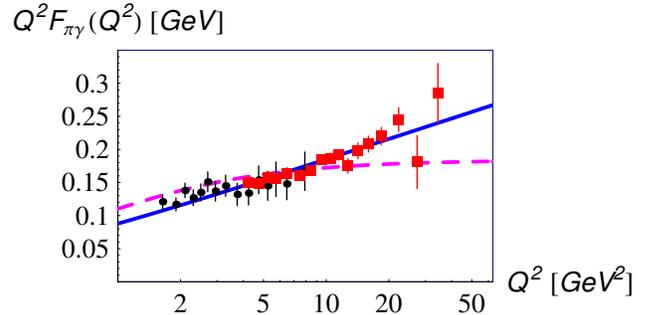}
\end{tabular}
\caption{\label{Plot:0}
Form factor $F_{\pi\gamma}$ vs $Q^2$: experimental data from \cite{cello,cleo} (black dots) and 
\cite{babar} (red squares);
dashed line represents the interpolation~\cite{bl} which coincides  
with the local-duality model of \cite{blm}, the solid line represents  
our fit described below.}
\end{center}
\end{figure}

The processes $\gamma^*\gamma \rightarrow \pi(\eta,\eta')$ are of  
great interest for our understanding of QCD and the meson structure. 
In particular, the  dependence on the  
space-like momentum $Q$ of the
incoming virtual photon has been the subject of detailed experimental  
\cite{cello,cleo,babar2,babar,babar1} and extensive
theoretical investigations (recent references see  
\cite{radyushkin,roberts,dorokhov,agaev,teryaev2,bt,kroll,mikhailov,blm,lcqm,czyz}).
QCD factorization predicts for large $Q^2$ the asymptotic behaviour  
\cite{bl} $Q^2F_{\pi\gamma}(Q^2)\to \sqrt{2}f_\pi$, with
$f_\pi= 0.131$ GeV denoting the pion decay constant. Similar relations  
follow for $\eta$ and $\eta'$ after 
taking particle mixing into account. Within errors, this saturation  
property is indeed observed for the $\eta$ and $\eta'$ form factors.
However, recent high-$Q^2$ data up to 35 GeV$^2$ indicate that  
$Q^2F_{\pi\gamma}(Q^2)$ does not saturate at large $Q^2$ but increases further. Fig.\ref{Plot:0}  
compares these data with the
theoretical formula obtained by Brodsky and Lepage \cite{bl} which  
interpolates the values
of $F_{\pi \gamma}(Q^2)$ at $Q^2=0$ given by the axial anomaly with  
the asymptotic form mentioned above.
So far, no compelling theoretical explanation of the qualitatively  
different behaviour of the $\pi\gamma$ form factor compared to
$\eta\gamma$ and $\eta'\gamma$ form factors has been found; as  
concluded in \cite{roberts,bt,mikhailov,blm} the behaviour of the
$\pi\gamma$ form factor is hard to explain in QCD.

The exact anomaly sum rule \cite{teryaev} offers an interesting  
possibility \cite{teryaev2} to consider the transition form factor
without directly referring to QCD factorization theorems. 
The aim of  
this letter is to address the $Q^2$-behaviour of the $P\gamma$ 
transition form factor from this perspective.

Our starting point for a treatment of the three processes is the  
investigation of the amplitude
\begin{eqnarray}
\langle 0| j_{\mu}^5|\gamma(q_2)\gamma^*(q_1)\rangle =e^2T_{\mu\alpha\beta}(p|q_1,q_2)~
\varepsilon^\alpha_1\varepsilon^\beta_2, \nonumber\\
\qquad p=q_1+q_2.
\end{eqnarray}
Here $\varepsilon_{1,2}$ denote the photon polarization vectors.
This amplitude is considered here for $q_1^2 = - Q^2$ and $q_2^2=0 $. 
Its general decomposition contains four invariant form  
factors, but for our purpose only the single form factor 
with the Lorentz structure proportional to $p_{\mu}$ is needed \cite{blm}
\begin{equation}
\label{F}
T_{\mu\alpha\beta}(p|q_1,q_2)= p_\mu \epsilon_{\alpha\beta q_1 q_2} iF(p^2,Q^2)+ \ldots
\end{equation}
The invariant amplitude $F(p^2,Q^2)$ may be written in terms of its  
spectral representations in $p^2$ at fixed $Q^2$:
\begin{eqnarray}
F(p^2,Q^2)=\frac{1}{\pi}\int\limits_{4m^2}^\infty\frac{ds}{s-p^2}\,\Delta(s,Q^2).
\end{eqnarray}
In perturbation theory one obtains the spectral density as an  
expansion in powers of $\alpha_s$:
\begin{eqnarray}
&&\Delta_{\rm QCD}(s,Q^2)=\nonumber\\
&&\qquad \Delta^{(0)}_{\rm QCD}(s,Q^2)+\frac{\alpha_s}
{\pi}\Delta^{(1)}_{\rm QCD}(s,Q^2)+O(\alpha_s^2),
\end{eqnarray}
where the lowest order contribution $\Delta^{(0)}_{\rm QCD}(s,Q^2)$ is  
obtained from the triangle diagram with the axial current and two vector currents in the vertices  
\cite{teryaev,ms,m}
\begin{eqnarray}
\label{1loop}
\Delta^{(0)}_{\rm QCD}&=& \frac{1}{2\pi}\frac{1}{(s+Q^2)^2}
\left[Q^2\,w+2m^2\log\left(\frac{1+w}{1-w}\right)\right],\nonumber\\
&&\quad w=\sqrt{1- 4 m^2/s}.
\end{eqnarray}
Here $m$ denotes the mass of the quark propagating in the loop.
The integral of $\Delta^{(0)}_{\rm QCD}(s,Q^2)$ from $s=4m^2$ to  
infinity has the remarkable property to be independent of $Q^2$ and $m$ \cite{teryaev} and gives the axial anomaly \cite{abj}
\begin{equation}
\label{anom}
\int\limits_{4m^2}^\infty ds \,\Delta^{(0)}_{\rm QCD}(s,Q^2) = \frac{1}{2 \pi}.
\end{equation}
The Adler-Bardeen theorem \cite{ab} states that radiative corrections  
to the anomaly vanish, requiring that
\begin{equation}
\label{an1}
\int\limits_{4m^2}^\infty ds \,\Delta^{(i)}_{\rm QCD}(s,Q^2) = 0,\quad i\ge 1.
\end{equation}
This relation however does not require that all higher-order spectral  
densities $\Delta^{(i)}_{\rm QCD}(s,Q^2)$ vanish; only the integrals should be zero.
The two-loop spectral density $\Delta^{(1)}_{\rm QCD}(s,Q^2)$ was  
found to be identically zero \cite{teryaev1,2loop}.
Higher-order spectral densities have not been calculated. However,  
arguments have been given that
all higher-order spectral densities may not vanish identically  
\cite{blm}.

Non perturbative QCD interactions will strongly influence $F(p^2,Q^2)$  
producing a meson pole (and, correspondingly, a (near) delta function of  
its absorptive part) and a hadron continuum. Nevertheless, the integral over the entire
absorptive part $\Delta(s,Q^2)$ remains unchanged. It still represents the anomaly:
\begin{equation}
\label{Aps1}
\int\limits_0^\infty ds \,\Delta(s,Q^2) = \int\limits_{4m^2}^\infty ds \,\Delta_{\rm QCD}(s,Q^2) = 
\frac{1}{2 \pi}.
\end{equation}
For the case of the isovector $\frac{\bar u u-\bar dd}{\sqrt2}$ axial  
current, the spectrum contains the $\pi^0$-meson. Thus, the absorptive part of $F(p^2,Q^2)$ has the form
\begin{eqnarray}
\label{Aps2}
&&\Delta(s,Q^2) =
\\ \nonumber
&& \pi  \delta (s - m_{\pi}^2) ~ \sqrt{2} f_{\pi}~F_{\pi
\gamma}(Q^2) + \theta (s - s_{\rm th})~\Delta^{I=1}_{\rm cont}(s,Q^2).
\end{eqnarray}
Here $\Delta^{I=1}_{\rm cont}(s,Q^2)$ denotes the hadron-continuum contribution in the isovector channel.
$F_{\pi\gamma}(Q^2)$ then takes the form
\begin{equation}
\label{pigamma}
F_{\pi\gamma}(Q^2) = \frac{1}{2 \sqrt{2}~ \pi^2 f_{\pi}}
\left[1- 2 \pi\int\limits_{s_{\rm th}}^\infty ds\;\Delta^{I=1}_{\rm cont}(s,Q^2)\right].
\end{equation}
For the $\eta\gamma$ and $\eta'\gamma$ form factors, one needs to consider the isoscalar currents 
$\bar q q=(\bar uu+\bar dd)/\sqrt{2}$ and $\bar s s$, separately:
\begin{eqnarray}
\label{e1e2}
F_{\bar q q}(Q^2)&=&\frac{1}{2 \sqrt{2}~ \pi^2 f_{q}}\left[1-2\pi \int
\limits_{s_{\rm th}}^\infty ds \,\Delta^{I=0}_{\rm cont}
(s,Q^2)\right],
\nonumber \\
F_{\bar s s}(Q^2)&=& \frac{1}{2 \sqrt{2}~ \pi^2 f_{s}}\left[1-2\pi \int
\limits_{s_{\rm th}}^\infty ds \,\Delta^{\bar ss}_{\rm cont}
(s,Q^2)\right].\nonumber \\
\end{eqnarray}
For each channel, the relevant threshold $s_{\rm th}$ should be used.
Taking $\eta-\eta'$ mixing \cite{anisovich,feldmann} into account and using 
the corresponding quark charges one finds
\begin{eqnarray}
\label{etaeta}
F_{\eta \gamma} (Q^2)& = &\frac{5}{3 \sqrt{2}}F_{\bar q q}(Q^2)~
\cos{ \phi} - \frac{1}{3}F_{\bar s s}(Q^2)~\sin{\phi}, \nonumber\\
F_{\eta'\gamma} (Q^2)& = &\frac{5}{3 \sqrt{2}}F_{\bar q q}(Q^2)~  
\sin{\phi} + \frac{1}{3}F_{\bar s s}(Q^2)~\cos{\phi}. \nonumber\\
\end{eqnarray}
The $\eta-\eta'$ mixing angle $\phi$ is known to be  $\phi \simeq 39^o
$; the decay constants are taken to
be $f_q= 1.07 f_{\pi}~,~f_s = 1.36 f_{\pi}$ \cite{feldmann}.

According to (\ref{pigamma}) and (\ref{e1e2}), the calculation of the  
$P\gamma$ form factors requires an Ansatz for the continuum spectral densities $\Delta_{\rm cont}
(s,Q^2)$  for all three cases.

At this place the duality concept may be used by replacing the  
integrand from threshold to infinity by the integrand obtained from  
perturbative QCD. For large values of $s$, above the resonance region,
the  spectral density is anyhow expected to be very well represented  
by perturbative QCD. At lower $s$, however, the continuum spectral  
density is strongly distorted by strong interactions and may not be  
fully accounted for by the rapid onset of the perturbative expression.  
Moreover, the thresholds should be in general replaced by effective  
thresholds. This holds in particular for the $\bar q q$ and $\bar s s$ dispersion 
representations which determine the $\eta$ and $\eta'$ form factors. Below  
the corresponding effective thresholds, the isoscalar interaction is very weak and is expected  
to be negligible.

A simple Ansatz for $\Delta(s,Q^2)$ which can to some extent parametrize these effects is to write
\begin{eqnarray}
\label{Ct}
&&\Delta_{\rm cont}(s,Q^2)=\theta(s - s_{\rm th})R(s)\Delta^{(0)}_{\rm QCD}(s,Q^2),\nonumber\\
&&\mbox{ with } R(s)=\left(1 - \frac{r}{s} \right).  
\end{eqnarray}
Finite value of $r$, $0\leq r \leq s_{\rm th}$, smoothens the threshold behavior. Moreover,  
the $r/s$ term strongly affects the high-$Q^2$ dependence of the form factor. 
We obtain this way a logarithmic increase with $Q^2$ of the form factor multiplied by $Q^2$: 
\begin{equation}
\label{FF}
Q^2 F( Q^2) \sim  \frac{Q^2}{Q^2+s_{th}} (s_{th}-r) + r\log\left(\frac{Q^2+s_{th}}{s_{th}}\right).
\end{equation}
To derive (\ref{FF}), we set $m=0$ (see the remark below). One may of course assume a more 
sophisticated behavior in the near-threshold region; in fact, the details of $\Delta_{\rm cont}(s,Q^2)$ at 
low $s$ influence the form factor at small $Q^2$. However, these details
do not change the $1/Q^2$-behaviour of $F_{P\gamma(Q^2)}$ at large $Q^2$.
Our crucial observation is that a wanted $\sim\log(Q^2)$ behavior of  
$Q^2 F_{P\gamma}(Q^2)$ requires the presence of the $r/s$ term in $R(s)$ (\ref{Ct}).

\begin{figure}[t!]
\begin{center}
\begin{tabular}{c}
\includegraphics[width=8.4cm]{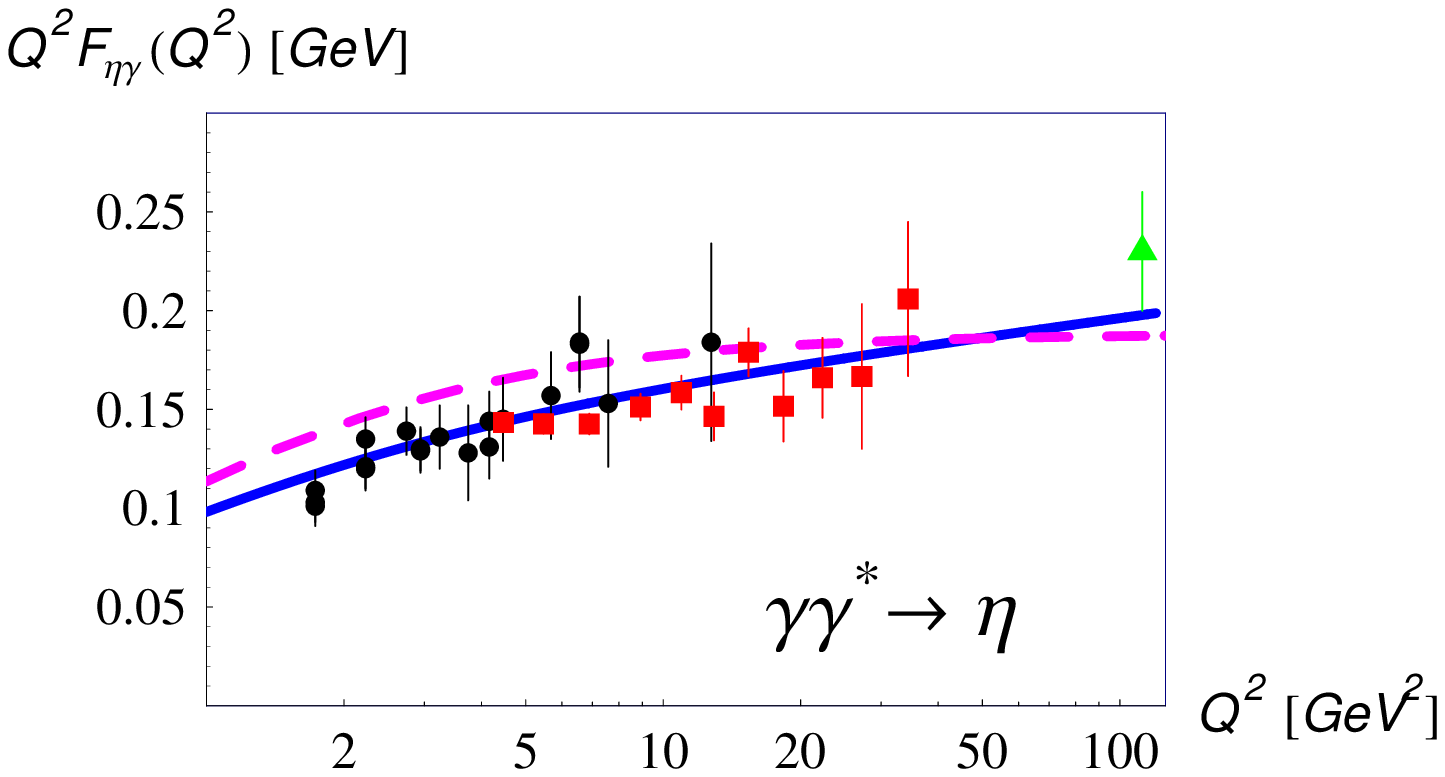}\\
\includegraphics[width=8.4cm]{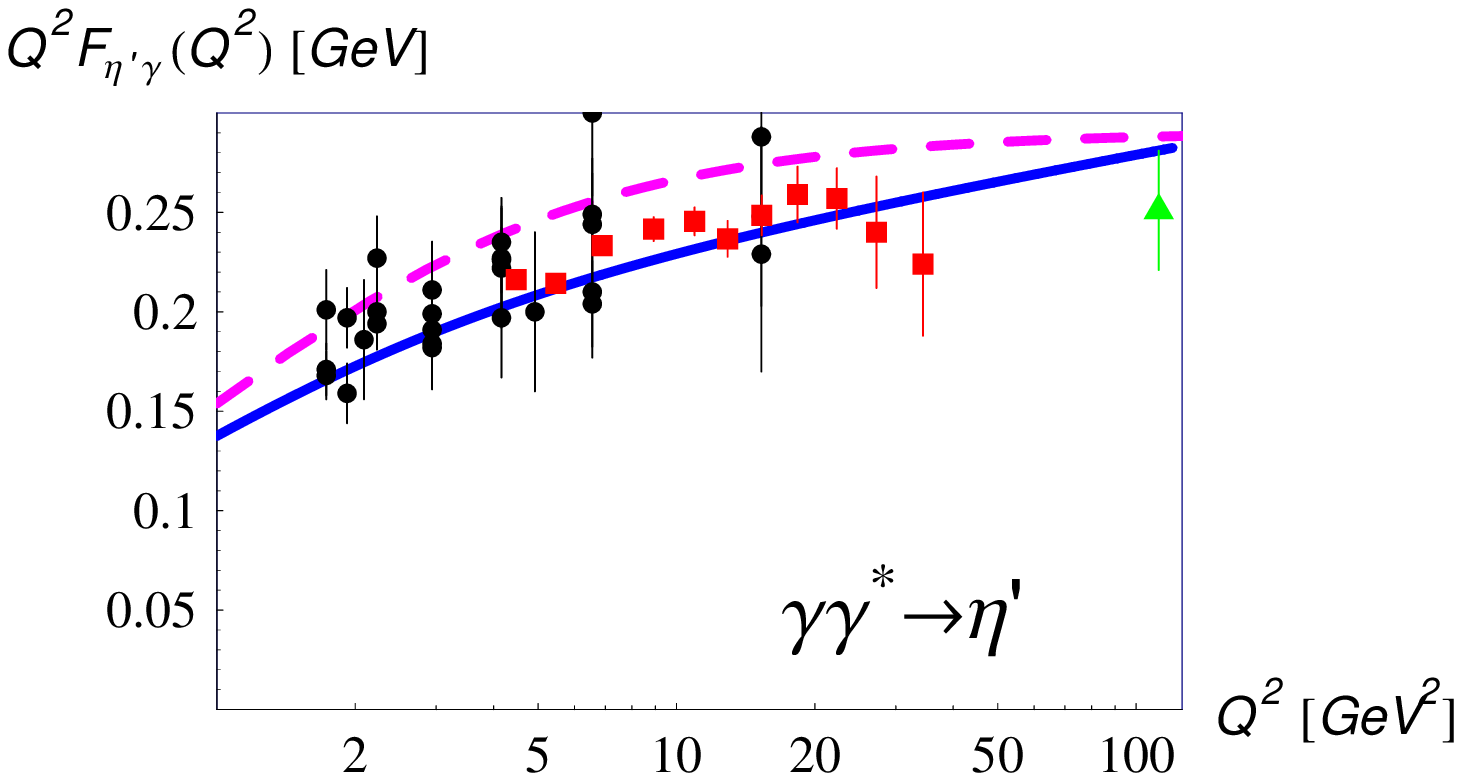}
\end{tabular}
\caption{\label{Plot:1}
Form factors $F_{P\gamma}$ ($P=\eta,\eta'$) vs $Q^2$: experimental data  
from \cite{cello,cleo} (black dots), \cite{babar1} (red squares), 
and the data borrowed from the time-like region \cite{babar2} (green triangle). 
Dashed line - results from local-duality model of \cite{blm}, solid line - our fit.}
\end{center}
\end{figure}

Making use of the formula (\ref{FF}), one finds values of the fit  
parameters $s_{\rm th}$ and $r$ which lead to a good
description of all existing experimental data, Figs.~\ref{Plot:0} and  
\ref{Plot:1}.
These parameters are as follows: for the pion case we use the physical threshold
value $s_{\rm th}=(3m_{\pi})^2=0.165$ GeV$^2$ and set also $r=(3m_{\pi})^2$;
then the continuum spectral density vanishes at $s_{\rm th}$.

It is less obvious which parameter values for the $\eta$ and $\eta'$  
cases should be used because their structure is influenced by the strong gluon anomaly \cite{feldmann}. 
We set $s_{\rm th} = 0.56$ GeV$^2$ for the $\bar qq$ channel and $s_{\rm th} = 0.76$ GeV$^2$ for 
the $\bar s s$ channel. 
These values are close to the $\{11\}$ and the $\{22\}$ elements of the $\eta - \eta'$ mass matrix, 
respectively \cite{feldmann}. 
The fit for the $r$ parameters yields $r=0.05$ GeV$^2$ for both channels. 
A good description of the data is achieved, see Fig.~\ref{Plot:1}. Notice that 
this value of $r$ is $\sim 1/3$ of the corresponding value in the $I=1$ channel. 
With regard to the quark mass values, one can safely set $m=0$ in (\ref{Ct}), at least for the $\pi$-meson case.
For the $F_{\bar q q}$ and $F_{\bar s s}$ amplitudes needed for the $\eta$ and $\eta'$ form factors the
dependence on possible values for $m_q$ and the strange mass value $m_s$ can easily be checked.
As long as one sticks to current quark masses there is no visible difference within present accuracies
from setting also for these cases $m_q = m_s = 0$. Therefore, the  
integrals involved above can be calculated analytically. The result has already been shown in (\ref{FF}).

\section{Discussion}
We studied the $\gamma^*\gamma\to P$ transition form factors by means  
of the exact anomaly sum rule which relates the integral over the hadron spectrum to 
the axial anomaly. In order to isolate the pseudoscalar mesons from this sum rule,
we make use of quark-hadron duality in the following way: we replace  
the integral over the continuum by the integral over the spectral density given by the diagrams of   
perturbation theory. We include, however---essentially,
by hand---an additional multiplication factor $R(s)=\left(1 - \frac{r}{s}
\right)$. This factor goes to unity for high values of $s$ above the resonance  
region where perturbative QCD should fully describe the continuum. This multiplication factor serves two  
purposes:
(a) It simulates a better threshold behavior and
(b) it modifies the high-$Q^2$ dependence of the integral over the  
continuum in a way suggested by recent high-$Q^2$
measurements. Our main results are as follows:

\vspace{.2cm}
\noindent
(i) The strong difference between the hadron spectral density and the  
perturbative QCD spectral density in the low-$s$ region influences only the low-$Q^2$ structure, 
but not the large-$Q^2$ asymptotics of the form factor
obtained from the anomaly sum rule. Using the parameter $r$ different from zero and  
positive, one easily obtains a good agreement for the low-$Q^2$ data, 
using threshold values well motivated by phenomenology. 
However, the increase observed for the $\pi\gamma$ form factor by BaBar cannot be understood
by subtleties of the behaviour of the continuum spectral density near  
threshold.

\vspace{.2cm}
\noindent
(ii) The deviation of the large-$Q^2$ behaviour of the form factor  
$F_{\pi\gamma}(Q^2)$ from the $1/Q^2$ asymptotics predicted by QCD factorization requires a  
deviation of order $O(r/s)$ between the hadron-continuum spectral density and the lowest-order  
perturbative QCD spectral density in the region of
larger $s$ values. The multiplication factor $R(s)$ in (\ref{Ct}) of the general form
\begin{eqnarray}
R(s)=1-\frac{r}{s}+\frac{r_1}{s^2}+\ldots
\end{eqnarray}
leads to the form factor which behaves at large $Q^2$ as
\begin{eqnarray}
&&Q^2F_{P\gamma}(Q^2)\sim  r\log(Q^2/s_{\rm th})\\
&&\qquad\qquad +\left(s_{\rm th}-r-r_1/s_{\rm th}\right)+O\left(\log(Q^2)/Q^2\right).\nonumber
\end{eqnarray}
The logarithmic increase of $Q^2F_{P\gamma}(Q^2)$ occurs proportional to $r$. 
Terms of higher order of $1/s$ in $R(s)$ do not contribute to the leading 
$\log(Q^2)/Q^2$-behaviour of 
the form factor.

\vspace{.2cm}
\noindent
(iii) The BaBar data on $F_{\pi\gamma}(Q^2)$ are well described by using $s_{\rm th}=(3m_\pi)^2$, 
the true threshold for starting the continuum integral, and by setting $r=(3m_\pi)^2$, a  
constant of the formal order
of the light-quark mass.
Since a good description of the data is obtained in this way, the  
theoretical treatment of the underlying $\langle AVV\rangle$
correlator in the chiral limit appears to be insufficient. Further  
studies of the effects beyond the chiral limit may be needed.   
Moreover, one should better understand the possible origin of such $O(r/s)$ terms within QCD. 
So far it is a mere conjecture suggested by the data.

\vspace{.2cm}
\noindent
(iv) Obviously, the proposed $O(r/s)$ term in $R(s)$ which modifies the perturbative QCD spectral density
leads to the violation of the QCD-factorization theorem. Only if the correction factor $R(s)$ is 
replaced by unity at some  
high value of $s$, the form factors multiplied by $Q^2$ will saturate  
at $Q^2$ larger than this $s$-value. Since we cannot say where this will  
occur - if it occurs at all - the true asymptotic behavior for very large $Q^2$ cannot be predicted.

\vspace{1ex}
\noindent{\it Acknowledgments.} We are grateful to W.~Lucha, J.~Pawlowski, and  
O.~Teryaev for valuable discussions.
D.~M.\ was supported by the Austrian Science Fund (FWF) under Project  
No.~P22843 and is grateful to the Alexander von
Humboldt-Stiftung and the Institute of Theoretical Physics of the
Heidelberg University for financial support and hospitality during
his stay in Heidelberg.


\end{document}